# XRL: An FMM-Accelerated SIE Simulator for Resistance and Inductance Extraction of Complicated 3-D Geometries

Mingyu Wang, Ping Liu, Jihong Gu, Xiaofan Jia, and Abdulkadir C. Yucel, *Senior Member, IEEE*

*Abstract*—**A fast multipole method (FMM)-accelerated surface integral equation (SIE) simulator, called XRL, is proposed for broadband resistance/inductance (RL) extraction under the magneto-quasi-static assumption. The proposed XRL has three key attributes that make it highly efficient and accurate for broadband RL extraction of complicated 3-D geometries: (i) The XRL leverages a novel centroid-midpoint basis transformation while discretizing surface currents, which allows converting edge-based vector potential computations to panel-based scalar potential computations. Such conversion makes the implementation of FMM straightforward and allows for drastically reducing the memory and computational time requirements of the simulator. (ii) The XRL employs a highly accurate equivalent surface impedance model that allows extracting RL parameters at low frequencies very accurately. (iii) The XRL makes use of a novel preconditioner, effectively including both diagonal entries and some near-field entries of the system matrix; such preconditioner significantly accelerates the iterative solution of SIE. The proposed XRL can accurately compute broadband RL parameters of arbitrarily shaped and large-scale structures on a desktop computer. It has been applied to RL parameter extraction of various practical structures, including two parallel square coils, a ball grid array (BGA) package and a high brand package on package. Its application to the parameter extraction of the BGA shows that the XRL requires 93.2x and 14.2x less computational time and memory resources compared to the commercial simulator Ansys Q3D for the same level of accuracy, respectively.**

## I. INTRODUCTION

T HE lumped circuit models realized by circuit parameters, such as resistance/inductance (RL) parameters under magneto-quasi-static (MQS) assumption, are of utmost significance for the power/signal integrity (PI/SI) analysis [1],

Manuscript received xxx. This work was supported by Xpeedic XPD.Hermes.023. (*Corresponding authors: Mingyu Wang, Abdulkadir C. Yucel.*)

M. Wang and P. Liu are with Xpeedic Co., Ltd. 60 NaXian Road, Building 5, Pudong New Area, 201210, Shanghai, China. (e-mails: mingyu.wang@xpeedic.com, ping.liu@xpeedic.com)

Jihong Gu is with the School of Microelectronics (School of Integrated Circuits), Nanjing University of Science and Technology, Xiaolingwei No.200, Nanjing, Jiangsu, China. (e-mail: gujihong@njust.edu.cn)

Xiaofan Jia and A. C. Yucel are with the School of Electrical and Electronic Engineering, Nanyang Technological University, Singapore 639798. (e-mail: xiaofan002@e.ntu.edu.sg, acyucel@ntu.edu.sg).

[2]. Such RL parameters can be extracted via various simulators, including mainstream volume integral equation (VIE) ones [3], [4], [5], [6], [7], [8], [9]. Among these, FastHenry [6] discretizes the volume currents within the structures via filaments and iteratively solves the VIE by performing fast multipole method (FMM)-accelerated matrix-vector multiplications (MVMs). However, such a filament-based discretization scheme becomes ineffective for complicated packaging structures with ground planes consisting of many vias and holes, rendering FastHenry inefficient. On the other hand, a tetrahedron-based discretization scheme, called TetraHenry [4], provides more flexible modeling of complicated 3D geometries. TetraHenry discretizes the volume currents via Schaubert–Wilton–Glisson basis functions [10] defined on tetrahedrons and leverages FMM to accelerate MVMs during the iterative solution of VIE. However, just like all volume discretization-based simulators, TetraHenry requires a large number of volume elements for the RL extraction at high frequencies due to the small skin depths. VoxHenry [3] is another open-source VIE simulator proposed for the RL parameter extraction of voxelized structures. It discretizes the volume currents with piecewise constant and linear basis functions and accelerates the MVMs via a fast Fourier transform (FFT). However, its uniform voxel discretization requirement makes it inefficient for RL extraction of complicated circuits or packages.

Besides VIE simulators, several surface integral equation (SIE) simulators have been developed for the RL parameter extraction [11], [12], [13], [14]. The SIE simulators often require far less number of elements to discretize the currents compared to VIE solvers and are thereby more efficient compared to VIE solvers. Among SIE solvers, open-source FastImp [11] supports MQS, electro-magneto-quasi-static, and full-wave analyses. It leverages a pre-corrected FFT (pFFT) scheme to accelerate MVMs, but requires a large number of unknowns (around seven times the number of panels) to be solved for MQS analysis. In addition, rectangle-based discretization is not highly suitable for complicated structures. Another powerful SIE technique explained in [12] converts the VIE formulation by introducing an equivalent surface impedance (ESI) model. Since its discretization uses the outer surfaces of filaments, its efficiency may suffer from the same limitations of filament-based FastHenry. Furthermore, the nodal analysis used in the study introduces extra potential unknowns, while the nodal incidence matrix worsen the condition number of the overall matrix system, which all reduces the efficiency of the simulator. Recently, a loop analysis-based SIE simulator for RL parameter extraction was proposed [14]. This simulator employs loop analysis to avoid the inclusion of extra potential unknowns. It uses Rao-Wilton-Glisson (RWG) basis function to discretize the surface currents and pFFT to accelerate MVMs. While it is efficient, its ESI model is not accurate enough in the low-frequency regime, when the skin depth is large, thereby the simulator cannot provide accurate broadband RL parameters (from very low to high frequencies). On the other hand, commercial full-wave simulation packages (Ansys HFSS and 3D-layout) call Ansys Q3D [15] for the enhanced accuracy and efficiency in the low-frequency regime [16]. Q3D RL engine employs an SIE simulator (AC RL engine) to extract the RL parameters in the high-frequency regime and calls a finite element method-based simulator (DC RL engine) with tetrahedral discretization to compute the RL parameters in the low-frequency regime [17]. The RL parameters in the transition frequency region are obtained by an approximate blending model [17], which introduces inaccuracy, especially when the interconnects' cross-sections change along the interconnects' longitudal directions. That's say, RL parameters obtained by Q3D in the transition region are not computed by an EM solver. To this end, to the best of our knowledge, there still exists a gap and demand for an accurate and efficient SIE simulator for broadband RL extraction from very low to high frequencies.

In this study, XRL is proposed to accurately and efficiently extract the broadband RL parameters of complicated 3-D structures. The XRL solves the SIE after discretizing the surface currents on triangular and rectangular supports via applying a novel centroid-midpoint (CM) basis transformation. After discretizing the currents and applying the Galerkin testing, the loop analysis is implemented to increase the efficiency of the simulator further. During the iterative solution of the linear system of equations (LSE), the proposed XRL accelerates the MVMs via FMM and drastically reduces the number of iterations with a specially developed preconditioner. The main contributions and unique features of the XRL are threefold:

1. The XRL utilizes a novel CM basis transformation while discretizing surface currents. Such transformation allows mapping the current vectors on edges to the panel centers. Doing so allows for converting the integrations on the traditional edge-based RWG/rooftop basis functions to those on the panel-based pulse basis functions. Such conversion reduces the computational requirements of the fast MVMs via FMM by a factor of 1.5~2. Furthermore, it drastically reduces the matrix-fill time for the near-field interaction, makes implementing matrix-fill routines straightforward, and yields remarkably less setup time for the preconditioner.

2. The XRL employs a highly accurate ESI model [12], [18], [19] that ensures the accuracy of RL extraction at the low-frequency regime, which makes the broadband RL extraction using a consistent algorithm, while commercial simulator Q3D requires two different solvers and one approximate blending model.

3. The XRL leverages a novel preconditioner, called diagonal-of-P preconditioner, which is built efficiently as the CM basis transformation is used (as indicated in the first feature of XRL). This novel diagonal-of-P preconditioner includes both the entries in the traditional diagonal-of-L preconditioner [20] and some near-field entries, obtained by multiplying the diagonal entries with sparse mapping matrices; such preconditioning scheme reduces the number of iterations significantly, as shown in the numerical results section.

The accuracy and efficiency of the proposed XRL simulator are demonstrated and compared with those of Ansys Q3D through their applications to the RL parameter extraction of various complicated 3-D structures, including parallel square coils, a ball grid array (BGA) package, and a high brand package on package (HBPOP). The numerical results show that the proposed XRL outperforms the Q3D on these practical examples. In particular, the XRL requires 4.38x less memory and 7.94x less CPU time compared to Q3D for a similar number of unknowns in the analysis of an HBPOP. Moreover, the XRL requires 14.2x less memory and 93.2x less CPU time compared to Q3D for the same level of accuracy in the analysis of a BGA. It should be noted that SIE solvers are capable of accurate RL extraction of general ICs, packages, and PCB structures. For very complicated structures, VIE solvers are often preferable due to their accuracy, versatility, and efficiency.

The rest of this paper is organized as follows. In Section II, the abovementioned unique features of XRL are explained in detail. In Section III, the numerical results demonstrating the applicability, accuracy, and efficiency of the XRL are presented. Finally, the paper is concluded with a discussion in Section IV.

## II. FORMULATION

In this section, the SIEs solved by XRL, their discretization by applying CM basis transformation, and the resulting LSE are explained first. Next, the details of the loop analysis, FMM acceleration, and the novel preconditioner are provided.

### A. SIEs, Their Discretization, and LSE

The proposed XRL simulator solves the SIE with the current continuity equation [12], which read

$$Z_s \mathbf{J}(\mathbf{r}) + \frac{j\omega\mu_0}{4\pi} \int_{S'} \frac{\mathbf{J}(\mathbf{r}')}{\|\mathbf{r} - \mathbf{r}'\|} d\mathbf{r}' = -\nabla\Phi \ , \qquad (1)$$

$$\nabla \cdot \mathbf{J}(\mathbf{r}) = 0 \ . \qquad (2)$$

Here $\omega = 2\pi f$ is the angular frequency, $f$ is the frequency, and $\mu_0$ denotes free space permeability. $\mathbf{r}$ and $\mathbf{r}'$ denote the observation and source locations on surface $S'$, comprised of conductors with conductivity $\sigma$. $Z_s$ is the ESI. In this study, for the panels included by rings [12], [18], the ESI is implemented as explained in [12], [18], otherwise, the Sonnet double-plane model [19] is implemented. The double-plane model reads as

$$Z_s = \frac{(1+j)R_{RF}\sqrt{f}}{1-e^{\frac{-(1+j)R_{RF}\sqrt{f}}{R_{DC}}}}, \qquad (3)$$

where $R_{DC} = \frac{2}{\sigma\varsigma}$, $R_{RF} = \sqrt{\frac{\pi\mu_0}{\sigma}}$, $\varsigma$ is the thickness of the conductor. In this study, the surface(s) of conductor(s) is/are discretized by the hybrid of triangles and rectangles, which provides a flexible meshing for complicated 3-D geometries. According to the PEEC model (see Fig. 1.), the centroids of adjacent panels $S_i$ and $S_j$ are regarded as circuit nodes and the edge $e$ shared by $S_i$ and $S_j$ is the circuit branch linking the two nodes. The global edge current vector $\mathbf{I}_e$ flows between the two nodes. Based on such a PEEC model, this study proposes a novel CM basis transformation applied while discretizing the surface current $\mathbf{J}$. To do that, the local current coefficient vector $\boldsymbol{\rho} = \mathbf{m}_e - \mathbf{c}_p$ is introduced where $\mathbf{m}_e$ and $\mathbf{c}_p$ denote the midpoint of an edge and centroid of a panel, respectively (see Fig. 2). Different from half-RWG basis function, the proposed CM basis transformation is free from quadrature points; it can be easily applied to arbitrary polygons and always satisfies the current continuity.

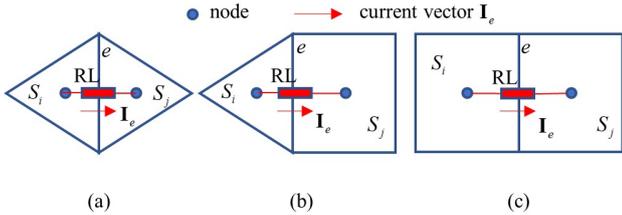

Fig. 1. The PEEC model in the adjacent two panels: (a) triangle and triangle (b) triangle and rectangle, and (c) rectangle and rectangle.

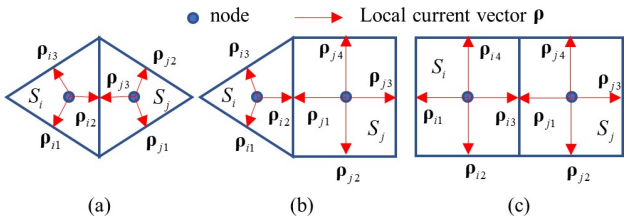

Fig. 2. The current coefficient vectors in the adjacent two panels: (a) triangle and triangle, (b) triangle and rectangle, and (c) rectangle and rectangle.

To map the current interactions to panels, three one-time generated sparse matrices $\overline{\mathbf{A}}_1 \in \mathbb{R}^{N_e \times N_p}$, $\overline{\mathbf{A}}_2 \in \mathbb{R}^{N_e \times N_p}$, and $\overline{\mathbf{A}}_3 \in \mathbb{R}^{N_e \times N_p}$ are introduced, where $N_e$ and $N_p$ are the number of edges and panels, respectively. The pseudocode provided in Algorithm 1 explains the procedure to obtain these sparse matrices. Note that these sparse matrices are generated during the loop search and do not require computational overhead.

---

**Algorithm 1** Procedure to generate $\overline{\mathbf{A}}_1$, $\overline{\mathbf{A}}_2$, and $\overline{\mathbf{A}}_3$

---

1 Obtain all local current coefficient vector $\boldsymbol{\rho}_{ik}$ and $\boldsymbol{\rho}_{jl}$ on panels $S_i$ and $S_j$, respectively. Here $k,l$ are the contour edge indices of panels $S_i$ and $S_j$, respectively.

2 If the direction of $\boldsymbol{\rho}_{ik}$ is the same as that of global edge current vector $\mathbf{I}_e$ in PEEC model (shown in Fig. 1.), $\overline{\mathbf{A}}_1(k,i) = \boldsymbol{\rho}_{ik,x}$, $\overline{\mathbf{A}}_2(k,i) = \boldsymbol{\rho}_{ik,y}$, and $\overline{\mathbf{A}}_3(k,i) = \boldsymbol{\rho}_{ik,z}$, otherwise, $\overline{\mathbf{A}}_1(k,i) = -\boldsymbol{\rho}_{ik,x}$, $\overline{\mathbf{A}}_2(k,i) = -\boldsymbol{\rho}_{ik,y}$, and $\overline{\mathbf{A}}_3(k,i) = -\boldsymbol{\rho}_{ik,z}$. Here $x,y,z$ are the $x$-, $y$-, and $z$-components of the local current coefficient vector.

3 Similarly, if the direction of $\boldsymbol{\rho}_{jl}$ is the same as that of the global edge current vector $\mathbf{I}_e$ in PEEC model (shown in Fig. 1.), $\overline{\mathbf{A}}_1(l,j) = \boldsymbol{\rho}_{jl,x}$, $\overline{\mathbf{A}}_2(l,j) = \boldsymbol{\rho}_{jl,y}$, and $\overline{\mathbf{A}}_3(l,j) = \boldsymbol{\rho}_{jl,z}$, otherwise, $\overline{\mathbf{A}}_1(l,j) = -\boldsymbol{\rho}_{jl,x}$, $\overline{\mathbf{A}}_2(l,j) = -\boldsymbol{\rho}_{jl,y}$, and $\overline{\mathbf{A}}_3(l,j) = -\boldsymbol{\rho}_{jl,z}$.

The surface current $\mathbf{J}$ is discretized via pulse basis functions after applying CM basis transformation as

$$\mathbf{J}(\mathbf{r}') = \sum_{e=1}^{N_e} \mathbf{I}_e = \sum_{l=1}^{N_p} w_l(\mathbf{r}')\overline{\mathbf{A}}_1(:,l)I_{p1}$$
$$+ \sum_{l=1}^{N_p} w_l(\mathbf{r}')\overline{\mathbf{A}}_2(:,l)I_{p2} + \sum_{l=1}^{N_p} w_l(\mathbf{r}')\overline{\mathbf{A}}_3(:,l)I_{p3} \qquad (4)$$

where $w_l(\mathbf{r}') = 1$ for $\mathbf{r}' \in S_l$; $w_l(\mathbf{r}') = 0$, otherwise. $S_l$ denotes the surface of $l^{\text{th}}$ panel, $l = 1,...,N_p$. $I_{p1}, I_{p2}, I_{p3}$ are the panel current coefficients. Substituting (4) into (1) and (2), applying Galerkin testing to the resulting equations with $w_k(\mathbf{r})$, $k = 1,...,N_p$ yields the LSE as

$$\overline{\mathbf{A}}_1\overline{\mathbf{Z}}\mathbf{I}_{p1} + \overline{\mathbf{A}}_2\overline{\mathbf{Z}}\mathbf{I}_{p2} + \overline{\mathbf{A}}_3\overline{\mathbf{Z}}\mathbf{I}_{p3}$$
$$= (\overline{\mathbf{A}}_1\overline{\mathbf{Z}}\overline{\mathbf{A}}_1^T + \overline{\mathbf{A}}_2\overline{\mathbf{Z}}\overline{\mathbf{A}}_2^T + \overline{\mathbf{A}}_3\overline{\mathbf{Z}}\overline{\mathbf{A}}_3^T)\mathbf{I}_{\text{edge}} = \mathbf{V}_{\text{edge}}. \qquad (5)$$

Here $\mathbf{I}_{pt} = \overline{\mathbf{A}}_t^T\mathbf{I}_{\text{edge}}$, $t = 1,2,3$. The entries $\overline{\mathbf{Z}}_{kl}$ in system matrix are

$$\overline{\mathbf{Z}}_{kl} = Z_s \int_{S_k} w_k(\mathbf{r})w_l(\mathbf{r}')dS$$
$$+ \frac{j\omega\mu_0}{4\pi}\int_{S_k}\int_{S_l} w_k(\mathbf{r})w_l(\mathbf{r}')\frac{1}{\|\mathbf{r}-\mathbf{r}'\|}dS'dS. \qquad (6)$$

$\mathbf{I}_{\text{edge}}$ is the unknown current vector and $\mathbf{V}_{\text{edge}}$ is the excitation vector whose entries $\left[\mathbf{V}_{\text{edge}}\right]_b = \varphi_i - \varphi_j$, where edge $b$ is the common edge shared by panels $S_i$ and $S_j$, and $\varphi_i, \varphi_j$ are the potential on nodes $i, j$, respectively. To solve (5), loop analysis [6], [14] is implemented in this project. It should be noted that the proposed method is different from the one proposed in [21], where the vertices and quadrature points on triangles are used to define the currents.

### B. Loop Analysis

To satisfy the current continuity in (2), nodal analysis enforcing Kirchoff's current law is adopted in [3], [12] via a sparse nodal incidence matrix, which introduces extra potential unknowns and increases the condition number of the whole system matrix. On the other hand, loop analysis enforcing Kirchoff's voltage law adopted in [4], [6], [14] introduces much less unknowns but requires additional time for loop search. Under the loop analysis, the LSE (5) is rewritten as

$$\bar{\mathbf{M}}(\bar{\mathbf{A}}_1\bar{\mathbf{Z}}\bar{\mathbf{A}}_1^T + \bar{\mathbf{A}}_2\bar{\mathbf{Z}}\bar{\mathbf{A}}_2^T + \bar{\mathbf{A}}_3\bar{\mathbf{Z}}\bar{\mathbf{A}}_3^T)\bar{\mathbf{M}}^T\mathbf{I}_l = \mathbf{V}_l. \qquad (7)$$

Here $\bar{\mathbf{M}} \in \mathbb{R}^{N_l \times N_e}$ is a sparse loop matrix and $N_l$ is the number of loops. $\mathbf{I}_{edge} = \bar{\mathbf{M}}^T\mathbf{I}_l$ and $\mathbf{V}_l = \bar{\mathbf{M}}\mathbf{V}_{edge}$ are the loop current and voltage vectors, respectively. If the loop $l$ includes edges $e$ and the direction of $\mathbf{I}_{edge}(e)$ is the same/opposite as that of loop $l$, then $\bar{\mathbf{M}}(l,e) = 1$ / $\bar{\mathbf{M}}(l,e) = -1$. For the detailed procedure for searching loops, please see [6], [14].

### C.  FMM Acceleration

FMM can be used to accelerate the MVMs during the iterative solution of (7). Traditionally, the current vector $\mathbf{I}_l$ can be decomposed into its three scalar components [4], [6] and the computational complexity of MVM scales with $O(N_e)$. In this study, the FMM is implemented more efficiently. To do that, (7) is divided into two parts

$$\bar{\mathbf{Z}}_s\mathbf{I}_l + \bar{\mathbf{M}}(\bar{\mathbf{A}}_1\bar{\mathbf{P}}\bar{\mathbf{A}}_1^T + \bar{\mathbf{A}}_2\bar{\mathbf{P}}\bar{\mathbf{A}}_2^T + \bar{\mathbf{A}}_3\bar{\mathbf{P}}\bar{\mathbf{A}}_3^T)\bar{\mathbf{M}}^T\mathbf{I}_l = \mathbf{V}_l. \qquad (8)$$

Here $\bar{\mathbf{Z}}_s$ is the surface impedance matrix under loop analysis and its multiplication with $\mathbf{I}_l$ is computationally cheap. The second part in (8) is rewritten as

$$\bar{\mathbf{M}}(\bar{\mathbf{A}}_1\bar{\mathbf{P}}\mathbf{I}_{p1} + \bar{\mathbf{A}}_2\bar{\mathbf{P}}\mathbf{I}_{p2} + \bar{\mathbf{A}}_3\bar{\mathbf{P}}\mathbf{I}_{p3}) = \mathbf{V}_l', \qquad (9)$$

where $\mathbf{I}_{pi} = \bar{\mathbf{A}}_i^T\bar{\mathbf{M}}^T\mathbf{I}_l$, $i = 1,2,3$. FMM is used to accelerate $\bar{\mathbf{P}}\mathbf{I}_{pi}$ directly and computational complexity of MVM scales with $O(N_p)$ in this case. In general, $N_e / N_p = 1.5 \sim 2$. To this end, the proposed FMM implementation on the centroids of panels introduces a computational saving by a factor of 1.5~2 compared to the traditional FMM implementation on edges.

In addition, the near-field computation of FMM in this implementation is also very efficient. In the implementation of FMM in [6] and [4] or pFFT in [14], to fill the near-field matrix, the edge-based vector potential is computed. The computational complexity of this operation is $O(N_p)$ with a multiplicative factor of 9 or 16 since there are 3 or 4 edges defined on a triangle or rectangle (on each panel), respectively. In our implementation, the core system matrix is $\bar{\mathbf{P}}$ directly computed on centroids of triangles or rectangles and the computational complexity of filling near-field matrix is $O(N_p)$. To this end, the computational time requirement of near-field computation is 9 or 16 times less than those in the traditional edge-based FMM implementation. It should be noted that the sole difference between our FMM implementation and the traditional FMM implementation is that while ours is operated on the panels, which naturally satisfies the format of particle simulation that is the original format of FMM [22], [23], the traditional one is implemented on the edges. For more information on the traditional FMM implementation, the reader is referred to [4] and [6].

### D.  Preconditioning

An efficient preconditioner is needed to reduce the number of iterations during the iterative solution of LSE. In the literature, the diagonal-of-L preconditioner, defined by $\bar{\mathbf{Z}}_s + j\omega\bar{\mathbf{L}}_d$, where $\bar{\mathbf{L}}_d$ is formed by the diagonal entries of edge-edge interaction matrix [20], has been a common choice and adopted in [4], [6], [14]. In our study, we map the edge interactions to panel interactions and obtain the dense matrix $\bar{\mathbf{P}}$. To this end, the diagonal-of-L preconditioner becomes ineffective for our study, as shown in the numerical results section.

Let $\bar{\mathbf{P}}_s$ denote a symmetric positive definite matrix with nonzero entries obtained from $\bar{\mathbf{P}}$. For any $\mathbf{x}$, let $\mathbf{y} = \bar{\mathbf{A}}_i^T\bar{\mathbf{M}}^T\mathbf{x}$. Then $\mathbf{x}^T(\bar{\mathbf{M}}\bar{\mathbf{A}}_i\bar{\mathbf{P}}\bar{\mathbf{A}}_i^T\bar{\mathbf{M}}^T)\mathbf{x} = \mathbf{y}^T\bar{\mathbf{P}}\mathbf{y} > 0$ since $\bar{\mathbf{P}}$ is positive definite. Similarly, $\mathbf{x}^T(\bar{\mathbf{M}}\bar{\mathbf{A}}_i\bar{\mathbf{P}}_s\bar{\mathbf{A}}_i^T\bar{\mathbf{M}}^T)\mathbf{x} = \mathbf{y}^T\bar{\mathbf{P}}_s\mathbf{y} > 0$. So the system $(\bar{\mathbf{M}}\bar{\mathbf{A}}_i\bar{\mathbf{P}}\bar{\mathbf{A}}_i^T\bar{\mathbf{M}}^T)(\bar{\mathbf{M}}\bar{\mathbf{A}}_i\bar{\mathbf{P}}_s\bar{\mathbf{A}}_i^T\bar{\mathbf{M}}^T)^{-1}$ has positive eigenvalues, which means that the condition number of the system $(\bar{\mathbf{M}}\bar{\mathbf{A}}_i\bar{\mathbf{P}}\bar{\mathbf{A}}_i^T\bar{\mathbf{M}}^T)(\bar{\mathbf{M}}\bar{\mathbf{A}}_i\bar{\mathbf{P}}_s\bar{\mathbf{A}}_i^T\bar{\mathbf{M}}^T)^{-1}$ is bounded independent of the matrices $\bar{\mathbf{M}}$ and $\bar{\mathbf{A}}_i$, $i = 1,2,3$. Let $\kappa(\bar{\mathbf{X}}) = \lambda_{\max}(\bar{\mathbf{X}}) / \lambda_{\min}(\bar{\mathbf{X}})$ be the condition number of the matrix $\bar{\mathbf{X}}$, where $\lambda_{\max}(\bar{\mathbf{X}})$ and $\lambda_{\min}(\bar{\mathbf{X}})$ are the maximum and minimum eigenvalues, respectively. Let $\bar{\mathbf{B}} = \bar{\mathbf{M}}\bar{\mathbf{A}}_i\bar{\mathbf{P}}\bar{\mathbf{A}}_i^T\bar{\mathbf{M}}^T$, and $\bar{\mathbf{C}} = \bar{\mathbf{M}}\bar{\mathbf{A}}_i\bar{\mathbf{P}}_s\bar{\mathbf{A}}_i^T\bar{\mathbf{M}}^T$. For matrix $\bar{\mathbf{B}}\bar{\mathbf{C}}^{-1}$, there are some $\mathbf{y}$ such that $\bar{\mathbf{B}}\bar{\mathbf{C}}^{-1}\mathbf{y} = \lambda_{\max}(\bar{\mathbf{B}}\bar{\mathbf{C}}^{-1})\mathbf{y}$. Then for $\mathbf{x} = \bar{\mathbf{C}}^{-1}\mathbf{y}$, the generalized eigenvalue problem $\bar{\mathbf{M}}\bar{\mathbf{A}}_i\bar{\mathbf{P}}\bar{\mathbf{A}}_i^T\bar{\mathbf{M}}^T\mathbf{x} = \lambda_{\max}(\bar{\mathbf{B}}\bar{\mathbf{C}}^{-1})\bar{\mathbf{M}}\bar{\mathbf{A}}_i\bar{\mathbf{P}}_s\bar{\mathbf{A}}_i^T\bar{\mathbf{M}}^T\mathbf{x}$ is satisfied. Therefore, $\lambda_{\max}(\bar{\mathbf{B}}\bar{\mathbf{C}}^{-1}) = (\mathbf{x}^T\bar{\mathbf{M}}\bar{\mathbf{A}}_i\bar{\mathbf{P}}\bar{\mathbf{A}}_i^T\bar{\mathbf{M}}^T\mathbf{x}) / (\mathbf{x}^T\bar{\mathbf{M}}\bar{\mathbf{A}}_i\bar{\mathbf{P}}_s\bar{\mathbf{A}}_i^T\bar{\mathbf{M}}^T\mathbf{x})$ [20] and so there is a vector $\mathbf{u} = \bar{\mathbf{A}}_i^T\bar{\mathbf{M}}^T\mathbf{x}$, $i = 1,2,3$, such that $\lambda_{\max}(\bar{\mathbf{B}}\bar{\mathbf{C}}^{-1}) \le \max_\mathbf{u}(\mathbf{u}^T\bar{\mathbf{P}}\mathbf{u} / \mathbf{u}^T\bar{\mathbf{P}}_s\mathbf{u})$ and thus $\kappa((\bar{\mathbf{M}}\bar{\mathbf{A}}_i\bar{\mathbf{P}}\bar{\mathbf{A}}_i^T\bar{\mathbf{M}}^T)(\bar{\mathbf{M}}\bar{\mathbf{A}}_i\bar{\mathbf{P}}_s\bar{\mathbf{A}}_i^T\bar{\mathbf{M}}^T)^{-1}) \le \kappa(\bar{\mathbf{P}}\bar{\mathbf{P}}_s)$. This shows that the positive definite $\bar{\mathbf{P}}_s$ should be a good choice to build a preconditioner and the sparsest approach would be to take the diagonal of $\bar{\mathbf{P}}$. Correspondingly, the diagonal-of-P preconditioner is proposed and employed for MQS analysis for the first time in this study. The proposed diagonal-of-P preconditioner is defined as

$$\bar{\mathbf{Q}} = \bar{\mathbf{Z}}_s + \bar{\mathbf{M}}(\bar{\mathbf{A}}_1\bar{\mathbf{P}}_d\bar{\mathbf{A}}_1^T + \bar{\mathbf{A}}_2\bar{\mathbf{P}}_d\bar{\mathbf{A}}_2^T + \bar{\mathbf{A}}_3\bar{\mathbf{P}}_d\bar{\mathbf{A}}_3^T)\bar{\mathbf{M}}^T, \qquad (10)$$

where $\bar{\mathbf{P}}_d$ is a diagonal matrix whose nonzero entries are the diagonal entries of $\bar{\mathbf{P}}$. Here the multiplication of $\bar{\mathbf{P}}_d$ with sparse mapping matrices $\bar{\mathbf{A}}_i$, $i = 1,2,3$, indicates that the self-interactions of edges and nearby interactions among all edges of each panel are included in the preconditioner, while the traditional diagonal-of-L preconditioner just includes the self-interactions of edges. During each MVM, the preconditioner is applied to (7) as

$$\mathbf{Q}^{-1}\overline{\mathbf{M}}(\overline{\mathbf{A}}_1\overline{\mathbf{Z}}\overline{\mathbf{A}}_1^T + \overline{\mathbf{A}}_2\overline{\mathbf{Z}}\overline{\mathbf{A}}_2^T + \overline{\mathbf{A}}_3\overline{\mathbf{Z}}\overline{\mathbf{A}}_3^T)\overline{\mathbf{M}}^T\mathbf{I}_l = \mathbf{Q}^{-1}\mathbf{V}_l. \quad (11)$$

The preconditioner-vector multiplication in (11)is achieved by Intel math kernel library (MKL) Pardiso LU decomposition. Note that for multi-frequency simulation, the ESI component is frequency-dependent and it will be computed for each frequency point. Fortunately, ESI component is computationally cheap to evaluate. For the potential part, the frequency-dependent coefficient $j\omega\mu_0/(4\pi)$ is extracted outside the integration, which means that the part is just computed for the first frequency point and then it is scaled by the frequency-dependent coefficient $j\omega\mu_0/(4\pi)$. Just like the advantage shown in the near field computation of FMM, such preconditioner implementation also reduces the time requirement by a factor of $9 \sim 16$ as it is operated on the panels (not edges).

## III. NUMERICAL RESULTS

This section presents several numerical examples that show the applicability, memory and CPU efficiency, and accuracy of the proposed XRL simulator. In the following analysis, when applicable, the results obtained by XRL are compared with those obtained by Q3D [15] and/or VoxHenry [3]. The discrepancy between results is quantified through $L^2$ norm difference, $diff$, which is defined as

$$diff = \sqrt{\sum\nolimits_{i=1}^{N_f} \left| F(f_i) - \tilde{F}(f_i) \right|^2 \bigg/ \sum\nolimits_{i=1}^{N_f} \left| \tilde{F}(f_i) \right|^2}, \quad (12)$$

where $N_f$ is total number of frequency points, $F(f_i)$, $\tilde{F}(f_i)$ denote the compared R or L parameter at a given frequency $f_i$, computed by XRL and analytical formula or Ansys Q3D or VoxHenry. The smallest skin depth during the given simulation is denoted as $\delta_s$. The proposed XRL simulator was implemented in C++. The LSE in (7) is iteratively solved by multiple right-hand-side generalized minimal residual method (mRHS-GMRES) [24] with a restart every 50 iterations until the relative residual error (RRE) reaches to $10^{-4}$ or $10^{-6}$ when the frequency is larger or smaller than 1 MHz, respectively. Both XRL and Q3D are executed on 32 threads of an Intel Xeon Gold 6252 CPU @2.1GHz with 1 TB RAM.

### A. Validation Examples

*1) Wire:* First, the frequency-dependent resistance and inductance of a 50 $\mu m$ -long wire [Fig. 3(a)] with a radius of 5 $\mu m$ are computed using XRL, VoxHenry, and Q3D. Obtained parameters are compared to those obtained by the analytical formula [25], considered as the reference, [Figs. 3(a)-(b)]. The frequency is swept from 1 KHz to 10 GHz ( $\delta_s = 0.652\ \mu m$ ). Table I lists the $L^2$ norm differences of the parameters obtained by the simulators. Clearly, due to XRL's accurate ESI, the resistance obtained by XRL perfectly matches with that obtained by the analytical formula. In addition, while the inductances obtained by all simulators are highly accurate with less than 1% relative difference, Q3D yields the smallest relative difference.

*2) Two Parallel Square Coils:* Next, the proposed XRL simulator, Ansys Q3D, and VoxHenry are used to obtain the RL parameters of two parallel 100 $\mu m$ × 100 $\mu m$ square coils with cross-sections of 5 $\mu m$ × 5 $\mu m$ (width x height) [Fig. 3(c)]. The edge-to-edge distance between two coils is 5 $\mu m$ and the spacing between source and sink in each coil is 2 $\mu m$. The RL parameters from 1 kHz to 100 GHz ( $\delta_s = 0.2062\ \mu m$ ) computed by the XRL and Ansys Q3D simulators are compared with those obtained by VoxHenry because the VIE solvers are considered the golden standard for the RL parameter extraction [3] and can accurately model the volume currents flowing in the channel formed due to skin depth. The self-resistance $R_{1,1}$, mutual resistance $R_{1,2}$, self-inductance $L_{1,1}$, and mutual inductance $L_{1,2}$ are plotted in Figs. 3(c)-(f), respectively. Clearly, self and mutual resistances and mutual inductances obtained by XRL are much closer to those of VoxHenry compared to Ansys 3D. In addition, the mutual resistance $R_{1,2}$ contributes slightly to loop inductance and Ansys Q3D provided a wrong trend when the frequency is larger than 100 MHz [Fig. 3(d)]. The $L^2$ norm difference, number of panels, CPU time and memory requirements of both XRL and Q3D are listed in Table II. Clearly, XRL requires 2.4x less memory and 7.6x less CPU time than Q3D for a similar number of panels, while the accuracy of R parameters ( $L^2$-norm difference) obtained by XRL is better than those obtained by Q3D due to XRL's accurate ESI model. The accuracy of self-inductance obtained by Q3D is slightly better than that obtained by XRL, while the accuracy of mutual inductance obtained by XRL is better.

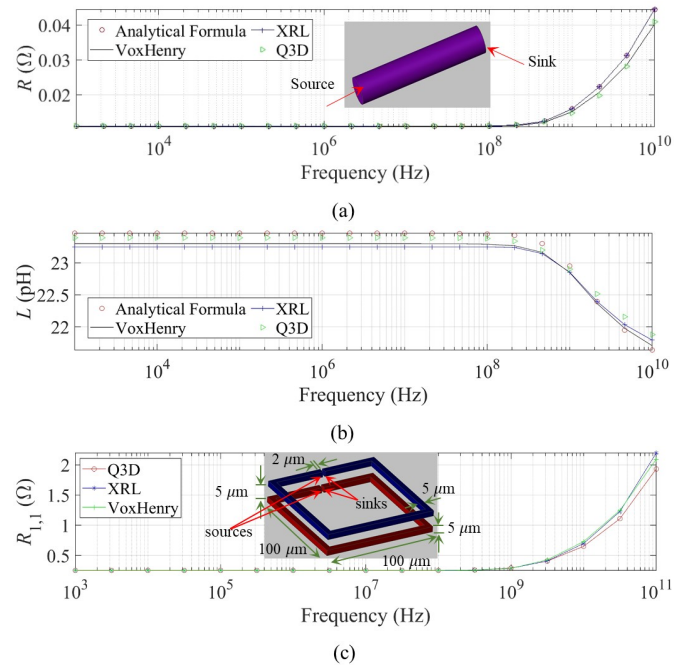

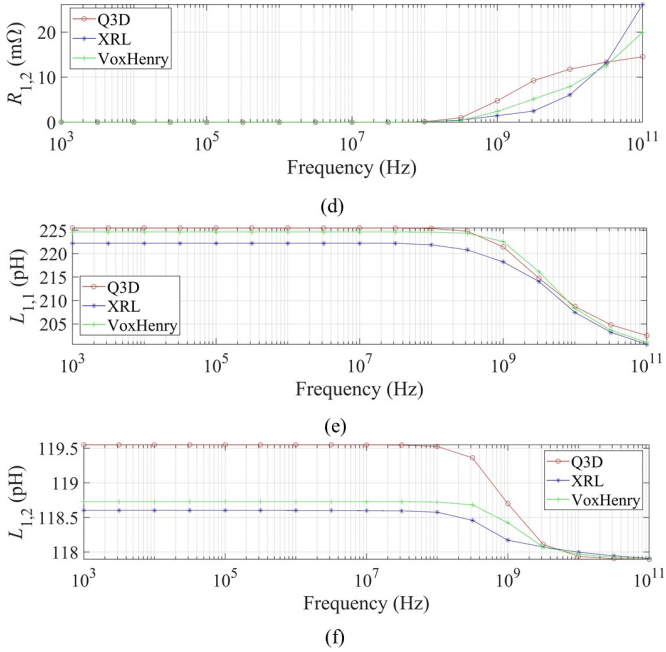

Fig. 3. Wire and parallel square coil examples. (a) The scenario of the wire and the comparisons of self-resistance and (b) self-inductance. (c) The scenario of the parallel square coils and the comparisons of self-resistance, (d) mutual resistance, (e) self-inductance, and (f) mutual inductance.

TABLE I
$L^2$ NORM DIFFERENCE FOR WIRE EXAMPLE

|  | *diff* in $R$ (%) | *diff* in $L$ (%) |
| --- | --- | --- |
| VoxHenry | 6.77 | 0.64 |
| Q3D | 6.98 | 0.43 |
| XRL | 0.27 | 0.83 |

TABLE II
NUMBER OF PANELS, MEMORY, AND CPU TIME REQUIREMENTS OF XRL AND ANSYS Q3D SIMULATORS FOR TWO PARALLEL SQUARE COILS EXAMPLE

|  | XRL | Ansys Q3D |
| --- | --- | --- |
| Number of Panels | 2188 | 2042 |
| Memory requirement (MB) | 78.3 | 189.6 |
| CPU time requirement (s) | 11 | 84 |
| *diff* in $R_{1,1}$ (%) | 4.03 | 8.28 |
| *diff* in $R_{1,2}$ (%) | 27.84 | 32.24 |
| *diff* in $L_{1,1}$ (%) | 1.11 | 0.42 |
| *diff* in $L_{1,2}$ (%) | 0.11 | 0.58 |

## B. A BGA Package

Next, a BGA package structure is considered [Fig. 4(a)]. Just like Ansys Q3D, XRL allows multiple source ports and only one sink port assigned on one signal net. In this structure, there are 25 signal nets and one ground net. On the terminals of wire bonds of signal nets, 50 source ports are defined. 25 sink ports are assigned on the bumps of all signal nets. In this example, the RL parameters obtained at 1.6 GHz ($\delta_s = 1.6299 \ \mu m$) via XRL are compared to those obtained by Ansys Q3D with 1% (the default setting) and 0.1% adaptive mesh refinement accuracy separately. Figs. 4(b)-(c) show that the self-inductances $L_{1,1}$, $L_{2,2}$, $\ldots$, $L_{50,50}$ of source ports $1,\ldots,50$ obtained by XRL match those obtained by Ansys Q3D with both 1% and 0.1% adaptive mesh refinement accuracy. However, the self-resistances $R_{1,1}$, $R_{2,2}$, $\ldots$, $R_{50,50}$ of source ports $1,\ldots,50$ obtained by XRL can only match those obtained by Ansys Q3D with 0.1% adaptive mesh refinement accuracy well. This is expected since the ESI model implemented in Q3D is not as accurate as that of XRL and directly affects the accuracy of the resistance extraction. Table III shows the memory and CPU time requirements of both simulators. For relatively the same level of accuracy, XRL requires 93x less CPU time and 14x less memory compared to Q3D.

In this example, the proposed diagonal-of-P preconditioner's performance is also tested. For this test, the tolerance of the iterative solver is set to $10^{-6}$. Table IV listed the setup time and the number of iterations required by the proposed diagonal-of-P preconditioner and traditional diagonal-of-L preconditioner, respectively. Needless to say, the proposed diagonal-of-P preconditioner performs better than the traditional diagonal-of-L preconditioner. Specifically, the proposed diagonal-of-P preconditioner requires 10.5x less setup time than the traditional diagonal-of-L preconditioner and the number of iterations is halved.

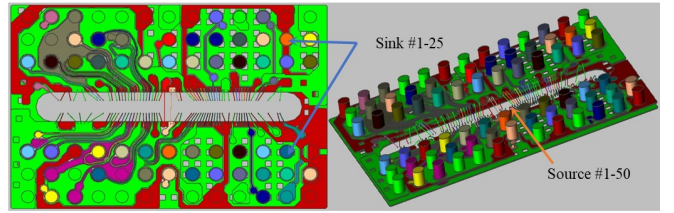

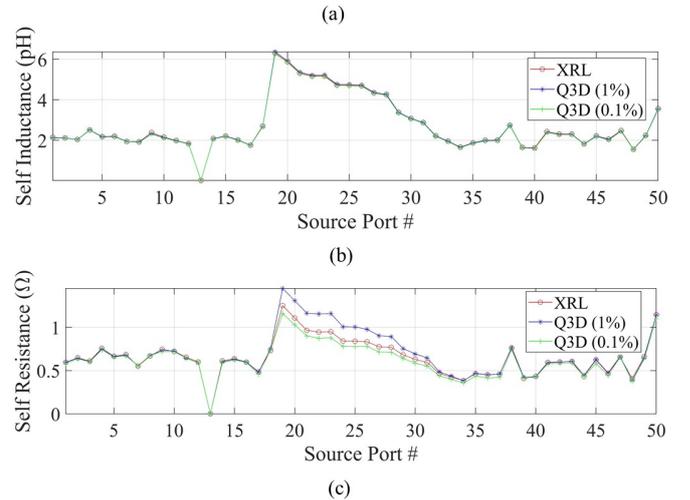

Fig. 4. A BGA package example. (a) The structure. (b) The self-inductances and (c) self-resistances of the source ports $1,\ldots,50$ at 1.6 GHz, corresponding to $L_{1,1},\ldots,L_{50,50}$ and $R_{1,1},\ldots,R_{50,50}$ are obtained by the XRL and Ansys Q3D.

TABLE III
MEMORY AND CPU TIME REQUIREMENTS OF XRL AND Q3D

|  | XRL | Q3D (1%) | Q3D (0.1%) |
| --- | --- | --- | --- |
| Memory requirement (GB) | 3.84 | 16.3 | 54.7 |
| CPU time requirement (h) | 0.155 | 3.05 | 14.45 |

TABLE IV
Setup Time and the Number of Iterations Achieved by the Proposed Diagonal-of-P Preconditioner and Traditional Diagonal-of-L Preconditioner

|  | Setup Time (s) | # Iterations |
|---|---|---|
| Diagonal-of-L preconditioner | 50.99 | 78 |
| Diagonal-of-P preconditioner | 4.85 | 36 |

### C. An HBPOP Package

Next, a realistic HBPOP package is considered [Fig. 5(a)]. The ports are added as shown in Fig. 5(a). The frequency is swept from 1Hz to 1GHz ( $\delta_s = 2.0617 \ \mu$m ). Figs. 5(b)-(c) show that $R_{1,1}$ and $R_{2,2}$ obtained by proposed XRL and Q3D; *diff* of $R_{1,1}$ and $R_{2,2}$ is computed as 1.94% and 1.89%, respectively. Figs. 5(d)-(f) show the self and mutual inductances obtained by XRL and Q3D. *diff* of $L_{1,1}$, $L_{1,2}$ and $L_{2,2}$ is computed as 3.53%, 3.90% and 3.47%, respectively. As shown in Table V, for relatively the same number of unknowns, XRL requires 7.94x less CPU time and 4.38x less memory compared to Q3D.

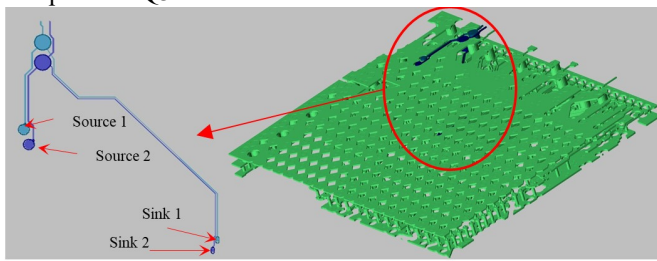

(a)

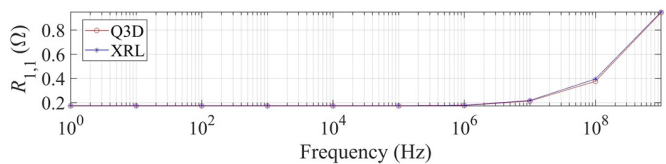

(b)

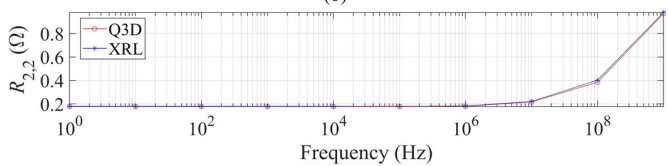

(c)

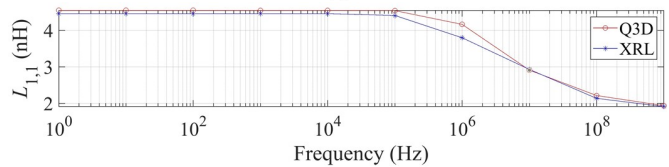

(d)

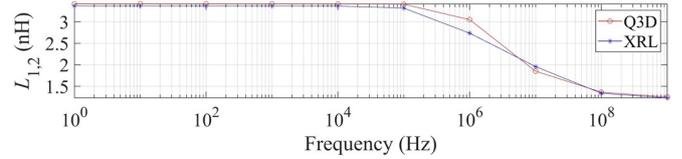

(e)

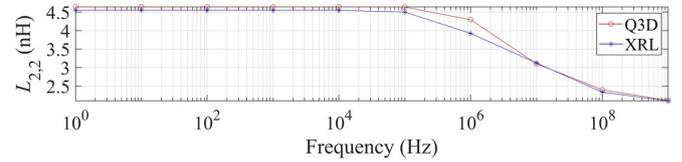

(f)

Fig. 5. An HBPOP package example. The comparison of the R and L parameters obtained by the XRL and Ansys Q3D. (a) The scenario of the FCBGA package and the comparisons of (b) $R_{1,1}$ , (c) $R_{2,2}$ , (d) $L_{1,1}$ , (e) $L_{1,2}$ , and (f) $L_{2,2}$ .

TABLE V
Memory and CPU Time Requirements of XRL and Q3D

|  | XRL | Q3D (0.1%) |
|---|---|---|
| # of Unknowns | 1231220 | 1219042 |
| Memory requirement (GB) | 47 | 206 |
| CPU time requirement (h) | 3 | 23.83 |

## IV. Conclusion

In this paper, XRL, an FMM-accelerated SIE simulator for broadband RL parameter extraction of complicated 3-D geometries, was introduced. The XRL solves SIEs after discretizing the surface currents with pulse basis functions and applying a novel CM basis transformation, which gives rise to simple implementation and a reduced computational cost. The proposed XRL simulator uses FMM to accelerate the MVMs during the iterative solution of LSE. Furthermore, it makes use of a highly effective and memory-efficient diagonal-of-P preconditioner to reduce the number of iterations. It leverages a highly accurate ESI model to achieve accurate results at low-frequency regime. For many practical scenarios, the proposed XRL simulator is much faster and memory efficient compared to the Ansys Q3D. For RL parameter extraction of two parallel square coils, the proposed XRL requires 2.4x less memory and 7.6x less CPU time compared to Q3D. Furthermore, for obtaining RL parameters of an HBPOP, the proposed XRL is more than 7.94x faster than Q3D, while it requires more than 4.38x less memory compared to Q3D for the same level of unknowns.

**Mingyu Wang** received the B.S. degree in electrical engineering and automation from Northeast Forestry University, China, in 2016, and M.S. and Ph.D. degrees in electronics from Nanyang Technological University, Singapore, in 2018 and 2022, respectively. He is currently working as an engineer in Xpeedic co., Ltd. to develop the RLCG parameters extraction solver.

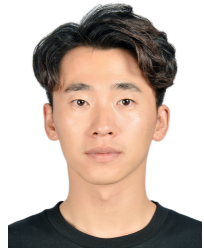

His research is focused on computational electromagnetics and fast parameter extraction.

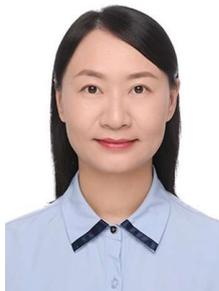

**Ping Liu** received the B.S. and M.S. degrees in communications engineering from Beijing Jiao Tong University, Beijing, China, in 2000 and 2003, respectively, and the Ph.D. degree in electromagnetic fields and microwave engineering from Shanghai Jiao Tong University, Shanghai, China, in 2006.

She was a R&D engineer in the Custom IC and PCB Group (CPG), Cadence Design Systems Inc., Shanghai, working on the model extraction and circuit simulation tools research and development in high-speed circuits and systems from 2006 to 2021. She joined Xpeedic Co., LTD, Shanghai, in September 2021 as a senior manager in algorithm division. Her current research interests include field solver modeling of high-speed interconnects, large-scale circuit system simulation, and power and signal integrity analysis in high-speed circuits.

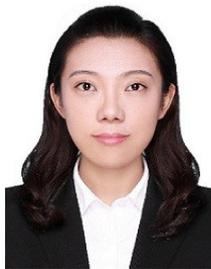

**Jihong Gu** (Member, IEEE) received the B.sc. degree in detection guidance and control technology and the Ph.D. degree in electronic information engineering from the School of Electrical Engineering and Optical Technique, Nanjing University of Science and Technology, Nanjing, China, in 2012 and 2019, respectively. She was a Research Scientist with the National University of Singapore, Singapore, from 2019 to 2022. Since 2022, she has been an Associate Professor with the School of Microelectronics (School of Integrated circuits), Nanjing University of Science and Technology. She is currently a Post-Doctoral Researcher with the Post-Doctoral Research Center of Changan Wangjiang, Chongqing, China. Her current research interests include computational electromagnetics, antenna, and RF-integrated circuits.

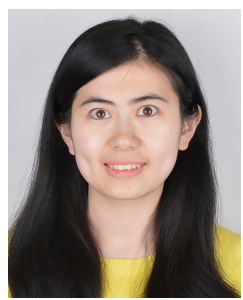

**Jia Xiaofan** received the B.Eng. degree from Nanjing University of Science and Technology, China in 2014 and the M.Sc. degree from Nanyang Technological University, Singapore in 2018. She is currently a Ph.D. candidate and a Research Associate at the School of EEE, Nanyang Technological University, Singapore. Her current research interests include computational electromagnetics, machine learning, and computer vision.

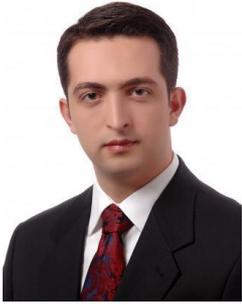

**Abdulkadir C. Yucel** (M'19-SM'20) received the B.S. degree in electronics engineering (Summa Cum Laude) from Gebze Institute of Technology, Kocaeli, Turkey, in 2005, and the M.S. and Ph.D. degrees in electrical engineering from the University of Michigan, Ann Arbor, MI, USA, in 2008 and 2013, respectively.

From September 2005 to August 2006, he worked as a Research and Teaching Assistant at Gebze Institute of Technology. From August 2006 to April 2013, he was a Graduate Student Research Assistant at the University of Michigan. Between May 2013 and December 2017, he worked as a Postdoctoral Research Fellow at various institutes, including the Massachusetts Institute of Technology. Since 2018, he has been working as an Assistant Professor at the School of Electrical and Electronic Engineering, Nanyang Technological University, Singapore.

Dr. Yucel received the Fulbright Fellowship in 2006, the Electrical Engineering and Computer Science Departmental Fellowship of the University of Michigan in 2007, and the Student Paper Competition Honorable Mention Award at IEEE AP-S in 2009. He has been serving as an Associate Editor for IEEE Antennas and Propagation Magazine, IEEE Journal on Multiscale and Multiphysics Computational Techniques, and International Journal of Numerical Modelling: Electronic Networks, Devices and Fields. His current research interests include various aspects of applied and computational electromagnetics with emphasis on ground penetrating, tree, and through-wall imaging radars, surrogate modelling and artificial intelligence techniques for expedient electromagnetic analysis, and fast integral equation-based frequency and time domain solvers with applications to the biomedical, circuit design, channel characterization in tunnels, and large-scale complex platforms.